\documentclass[11pt]{article}
\def\bc{\begin{center}}
\def\ec{\end{center}}
\def\beq{\begin{equation}}
\def\eeq{\end{equation}}

\def\pd{\partial}
\title{\bf Pomeron with a running coupling constant in the nucleus}
\author{M.A.Braun  \\
S.Peterburg State University, Russia
}
\begin{document}
\maketitle

\begin{abstract}
The running coupling is introduced into the equation for
propagation of the pomeron in the nucleus via the bootstrap relation.
The resulting equation coincides with the one obtained in the
colour dipole formalism by summing contributions from   quark-antiquark
loops, with a general choice of the regularization scheme.
\end{abstract}

\section{Introduction}
Lately a renewed interest has been shown towards introduction of a running
coupling into the BFKL dynamics.
Summation of contributions from the quark-
antiquark loops to the evolution of the gluon density has been used to
restore the full dependence of the coupling on the running scale in
the color dipole approach ~\cite{kovchegov1, kovchegov2, balitsky}.
It turned out that the obtained kernel for the linear BFKL equation
essentially
coincides with the one which we found many years ago by imposing the
bootstrap condition
necessary for the fulfilment of unitarity  ~\cite{braun1,braun2}.
In this paper we draw
attention to the fact that the bootstrap condition in fact allows to derive
also the
structure of the triple pomeron interaction and thus the form of the
non-linear
BFKL equation describing propagation of the pomeron in the nucleus.
This is a simple consequence of the possibility to express the basic
splitting kernel
from 2 to 3 and 4 gluons ('the Bartels vertex', ~\cite{bartels0})
via the basic BFKL interaction.
The relation between the splitting vertex and  reggeized gluon interaction,
together with the bootstrap, are  the necessary ingredients of the pomeron
interaction
in the reggeized gluon approach. Indeed they allow to present all
contributions to
the 4-gluon amplitude in the standard form of a pomeron splitting into
two pomerons.
Thus preserving these two relations seems to be essential for the
construction
of pomeron interaction with the running coupling.

As in our earlier paper, we have to stress from the start that introduction
of the running coupling into the BFKL formalism cannot be made rigorously
and uniquely.
The formalism admits transverse momenta of any magnitude, including very
small ones,
at which the concept of the gluon and its coupling looses any  sense.
The introduced
running coupling has to be artificially continued to small momentum
values,
where it is completely undetermined. The only information one can get from
the running of the coupling refers to the region of high momenta.
It remains to be
seen in what measure this information depends on the low momentum behaviour,
which
cannot be fixed theoretically in any reliable manner. It is known however
that
the non-linear BFKL equation, unlike the linear one, is not very sensitive
to the
infrared region. So hopefully this equation has a better chance to produce
results
weakly dependent on the assumed infrared behaviour of the running coupling.

The paper is organized as follows. The first section is dedicated to the
derivation of the triple-pomeron vertex with the running coupling
introduced via the bootstrap relation. After recalling this method to
introduce the running coupling, we go along the same steps as in
~\cite{braun3, bravac}, where the vertex was derived
in the limit $N_c\to\infty$ for the fixed coupling.
In Section 3 we construct the full amplitude with a single triple pomeron
interaction, coupling the vertex with three pomerons. This result
enables us to build the equation for the pomeron in the large nucleus in
Section 4. Finally in Section 5 we compare our results with those obtained
within   the colour dipole approach.

Having in mind application of the formalism to the large nucleus as a target,
we limit ourselves to the case of propagation of pomerons
with zero total momenta. However  generalization to non-zero total momenta is
straightforward.

\section{Triple-Pomeron vertex}
\subsection{Generalities}
As mentioned in the Introduction, this paper follows the idea to introduce
a running coupling via the bootstrap ~\cite{braun1,braun2}. Derivation
of the triple-pomeron vertex in the limit $N_c\to\infty$ then goes as
presented in ~\cite{braun3, bravac} for the fixed coupling case.

Basic formulas for the introduction of a running coupling via the bootstrap
condition consist in expressing both the gluon trajectory $\omega$ and
intergluon interaction in the vacuum channel $K$ in terms of a
single function $\eta(q)$ of the gluon momentum, which then can be chosen
to conform to the high-momentum behaviour of the gluon density with a
running coupling. Explicitly
\beq
\omega(q)=-\frac{1}{2}N_c\int \frac{d^2q_1}{(2\pi)^2}\frac{\eta(q)}
{\eta(q_1)\eta(q_2)},
\label{traj}
\eeq
\beq
K(q_1,q_2|q'_1q'_2)=N_c\Big[\Big(\frac{\eta(q_1)}{\eta(q'_1)}+
\frac{\eta(q_2)}{\eta(q'_2)}\Big)\frac{1}{\eta(q_1-q'_1)}-
\frac{\eta(q_1+q_2)}{\eta(q'_1)\eta(q'_2)}\Big].
\label{int}
\eeq
In these definitions it is assumed that $q_1+q_2=q'_1+q'_2=q$.
For arbitrary $\eta(q)$ the following bootstrap relation is satisfied:
\beq
\frac{1}{2}\int\frac{d^2q'_1}{(2\pi)^2}K(q,q_1,q'_1)=\omega(q)-\omega(q_1)-
\omega(q_2).
\eeq
The fixed coupling corresponds to the choice
\beq
\eta(q)=\frac{2\pi}{g^2}q^2.
\label{etafix}
\eeq
Then one finds the standard expression for the trajectory $\omega(q)$
and
\[K(q,q_1,q'_1)=\frac{g^2N_c}{2\pi}V(q,q_1,q'_1),\]
where $V$ is the standard BFKL interaction. Note that the extra $2\pi$
in the denominator corresponds to the  BFKL weight $1/(2\pi)^3$ in the
momentum integration, which we prefer to take standardly as $1/(2\pi)^2$.

From the high-momentum behaviour of the gluon distribution with a running
coupling one finds
\beq
\eta(q)=\frac{1}{2\pi}bq^2\ln\frac{q^2}{\Lambda^2},\ \ q^2>>\Lambda^2,
\label{asym}
\eeq
where $\Lambda$ is the standard QCD parameter and
\beq
b=\frac{1}{12}(11 N_c-2N_f).
\label{bval}
\eeq
As to the behaviour of $\eta(q)$ at small momenta, we shall assume
\beq
\eta(0)=0,
\label{eta0}
\eeq
which guarantees that the gluon trajectory $\omega(q)$
passes through zero at $q=0$ in accordance with the gluon properties.
The asymptotic (\ref{asym}) and condition (\ref{eta0}) are the only
properties of $\eta(q)$ which follow from the theoretical reasoning.
A concrete form of $\eta(q)$ interpolating between (\ref{eta0}) and
(\ref{asym}) may be chosen differently. One hopes that the following
physical results will not strongly depend on the choice.

Our old derivation in ~\cite{braun3} of the triple-pomeron vertex
was actually based on the property (\ref{eta0}) obviously valid for
(\ref{etafix}), the
bootstrap relation and the relation between the Bartels transition
vertex $K_{2\to 3}$ and intergluon BFKL interaction $V$  (Eq. (12) in
~\cite{braun3})
\beq
K_{2\to 3}(1,2,3|1',3')=V(2,3|1'-1,3')-V(12,3|1'3').
\label{oldw}
\eeq
Here and in the following we frequently denote gluon momenta just by numbers:
$1\equiv q_1$, $1'\equiv q'_1$ etc. Also we use $12\equiv q_1+q_2$.
All the rest conclusions were obtained from these three relations in a
purely algebraic manner.

Our idea is that if we {\it define} the transition vertex by a similar
relation in terms of  the new intergluon vertex $K$, Eq. (\ref{int}),
then all the derivation will remain valid also for arbitrary $\eta(q)$
satisfying (\ref{eta0})
and thus for a running coupling, provided $\eta(q)$ is chosen appropriately.
In the next subsection we briefly recapitulate successive stages of the
derivation for arbitrary $\eta(q)$ with $\eta(0)=0$.

\subsection{Leading order in $N_c\to\infty$}
The changes necessary to pass to arbitrary $\eta(q)$ are minimal.
Obviously in our old formulas we have to drop the coupling constant $g$
factors, since now $g$ is provided by $1/\eta$. We also drop factors $N_c$
in the interaction because we prefer to include it into the kernel $K$.
With these comments, the two-gluon equation becomes
\beq
S_{20}D_2=D_{20}+K_{12}D_2,
\label{eq2}
\eeq
where
\beq
S_{20}=j-1-\omega(1)-\omega(2),
\label{kin2}
\eeq
$j$ is the angular momentum variable,
\beq
D_{20}=N_c\Big(f(0,q)-f(q_1,q_2)\Big)
\eeq
and $f(q_1,q_2)$ is just the quark-antiquark loop with, say, the gluon
with momentum $q_1$ coupled to the quark and the other, with momentum $q_2$,
coupled to the antiquark.
With the running coupling $D_{20}$ does not change its form, but both
$\omega$ in (\ref{kin2}) and $K$ in (\ref{eq2}) are now given by
Eqs. (\ref{traj}) and (\ref{int}) respectively.

The three-gluon system exists in two colour states, which differ in the
ordering of the three gluons along the loop, 123 and 213.
For each order the equation is
\beq
S_{30}D_3=D_{30}+D_{2\to 3}+\frac{1}{2}(K_{12}+K_{23}+K_{31})D_3,
\label{eq3}
\eeq
where now
\beq
S_{30}=j-1-\sum_{j=1}^3\omega(j)
\eeq
and
\beq
D_{30}^{(123)}=-D_{30}^{(213)}=\sqrt{\frac{N_c}{8}}
\Big(D_{20}(2)-D_{20}(1)-D_{20}(3)\Big).
\label{ini3}
\eeq
The new element is the term $D_{2\to 3}$ which corresponds to transitions
of the initial two-gluon system into the final three-gluon system.
This transition is accomplished by the $2\to 3$ vertex $W$ which, as
explained, we define via the new intergluon interaction by a relation
similar to (\ref{oldw})
\beq
W(1,2,3|1',3')=K(2,3|1'-1,3')-K(12,3|1'3').
\label{neww}
\eeq
Explicitly, in terms of $\eta$
\beq
\frac{1}{N_c}W(1,2,3|1',3')=\frac{\eta(2)}{\eta(1-1')\eta(3-3')}
-\frac{\eta(23)}{\eta(1-1')\eta(3')}-\frac{\eta(12)}{\eta(1')\eta(3-3')}
+\frac{\eta(123)}{\eta(1')\eta(3')}.
\label{wexpl}
\eeq
Note that, as with a fixed coupling constant,
\beq
W(1,2,3|1',3')=W(3,2,1|3',1').
\eeq

We find in full analogy with ~\cite{braun3}
\beq
D_{2\to 3}=\sqrt{\frac{N_c}{8}}W(1,2,3|1',3')\otimes D_2(1',3')
\equiv \sqrt{\frac{N_c}{8}}W(1,2,3),
\label{d23}
\eeq
where $\otimes$ means integration over the intermediate gluon momentum with
weight $1/(2\pi)^2$.

Next step is to show that Eq. (\ref{eq3}) for the three-gluon system is
solved by the reggeized zero term ansatz:
\beq
D_{3}^{(123)}=-D_{3}^{(213)}=\sqrt{\frac{N_c}{8}}
\Big(D_{2}(2)-D_{2}(1)-D_{2}(3)\Big),
\label{sol3}
\eeq
that is by (\ref{ini3}) in which the loops are substituted by the full
solutions of the two-gluon equation (\ref{eq2}).
The proof is wholly based on the bootstrap and relation (\ref{neww})
and literally repeats the corresponding proof in ~\cite{braun3}.

Passing to the 4-gluon system, in the limit $N_c\to\infty$ we find two
configurations differing by  the order  of gluons along the quark-antuquark
loop: 1234 and 2134.
The equation governing the 4-gluon system is
\beq
S_{40}D_4=D_{40}+D_{2\to 4}+D_{3\to 4}+\frac{1}{2}(K_{12}+K_{23}+K_{34}+
K_{41})D_4,
\label{eq4}
\eeq
where
\beq
S_{40}=j-1-\sum_{j=1}^4\omega(j).
\eeq
The inhomogeneous terms are
\beq
D_{40}^{(1234)}=\frac{1}{4}N_c
\Big(D_{20}(1)+D_{20}(4)-D_{20}(14)\Big),
\label{ini41}
\eeq
\beq
D_{40}^{(2134)}=\frac{1}{4}N_c
\Big(D_{20}(2)+D_{20}(3)-D_{20}(12)-D_{20}(13)\Big),
\label{ini42}
\eeq
\beq
D_{2\to 4}^{(1234)}=-\frac{1}{4}N_cW(1,23,4),\ \
D_{2\to 4}^{(2134)}=0
\eeq
(the definition of $W(1,2,3)$ is given by the 2nd equality in (\ref{d23})),
\beq
D_{3\to 4}^{(1234)}=\sqrt{\frac{N_c}{8}}
\Big(W(2,3,4|2',4')\otimes D_3^{(124)}(1,2',4')+
W(1,2,3|1',3')\otimes D_3^{(134)}(1',3',4)\Big),
\eeq
and
\beq
D_{3\to 4}^{(2134)}=-\sqrt{\frac{N_c}{8}}
\Big(W(1,2,4|1',4')\otimes D_3^{(134)}(1',3,4')+
W(1,3,4|1',4')\otimes D_3^{(124)}(1',2,4')\Big).
\eeq

Repeating the corresponding derivation in ~\cite{braun3} we
demonstrate that in the limit $N_c\to\infty$
the solution of the 4-gluon equation is again given by the reggeized
zero-order terms:
\beq
D_{4}^{(1234)}=\frac{1}{4}N_c
\Big(D_{2}(1)+D_{2}(4)-D_{2}(14)\Big)
\label{sol41}
\eeq
and
\beq
D_{4}^{(2134)}=\frac{1}{4}N_c
\Big(D_{2}(2)+D_{2}(3)-D_{2}(12)-D_{2}(13)\Big).
\label{sol42}
\eeq
The proof is purely algebraic and is wholly based on the bootstrap and
relations (\ref{neww}) and (\ref{eta0}) valid for any choice of function $\eta(q)$
with $\eta(0)=0$.

\subsection{The triple pomeron configuration}
Next step is to study the next-to-leading configuration in $N_c\to\infty$
corresponding to the triple pomeron interaction.
Again the derivation practically literally repeats our old one for
a fixed coupling constant.
The governing 4-gluon equation is similar to Eq. (\ref{eq4})
\beq
S_{40}D_4=D_{40}+D_{2\to 4}+D_{3\to 4}+D_{4\to 4}+
(K_{12}+K_{34})D_4.
\label{eq40}
\eeq
The formal difference is in the absence of interaction between the two
final pomerons, which are assumed to be made of gluon pairs (1,2) and (3,4),
doubling of the rest interactions acting in the vacuum colour channels
and in the appearance of the term $D_{4\to 4}$, which describes transitions
from the leading colour configuration to the subleading one corresponding
to two pomerons.

The four inhomogeneous terms are
\beq
D_{40}=\frac{1}{2}\Big(\sum_{j=1}^4D_{20}(j)-\sum_{j=2}^4D_{20}(1j)\Big),
\eeq
\beq
D_{2\to 4}=-W(1,23,4),
\eeq
\[
D_{3\to 4}=\sqrt{\frac{2}{N_c}}\Big(
W(1,2,3|1',3')\otimes D_3^{(134)}(1',3',4)-
W(1,2,4|1',4')\otimes D_3^{(134)}(1',3,4')
\]\beq+
W(2,3,4|2',4')\otimes D_3^{(124)}(1,2',4)-
W(1,3,4|1',4')\otimes D_3^{(124)}(1',2,4)\Big)
\eeq
and
\beq
D_{4\to 4}=\frac{1}{N_c}(K_{23}+K_{14}-K_{13}-K_{24})(D_4^{(1234)}-
D_4^{(2134)}),
\eeq
where the $D_4$'s on the right-hand side are given by Eqs. (\ref{sol41})
and (\ref{sol42}).
As with a fixed coupling constant, all terms except $D_{40}$ can be presented
as a result of action of a certain operator $Z$ on the two gluon state,
so that Eq. (\ref{eq40}) can be rewritten as
\beq
S_{40}D_4=D_{40}+Z\otimes D_2+
(K_{12}+K_{34})D_4.
\label{eq41}
\eeq
The explicit form of this operator can be expressed in terms of a
function
\beq
G(1,2,3)=-W(1,2,3)-
D_2(1,23)(\omega(2)-\omega(23))-D_2(12,3)(\omega(2)-\omega(12)).
\label{defg}
\eeq
Then one finds
\[
Z\otimes D_2=\frac{1}{2}\Big(2G(1,34,2)+2G(3,12,4)+G(1,23,4)+G(1,24,3)+
G(2,13,4)+G(2,14,3)
\]\[
-G(1,3,24)-G(1,4,23)-G(2,3,14)-G(2,4,13)-G(3,2,14)\]\beq
-G(3,1,24)-G(4,2,13)-G(4,1,23)+G(23,0,14)+G(13,0,24)\Big).
\eeq
This formula  is identical to the old one in ~\cite{braun3}, but with
new expressions for both $W$ and $\omega$ in terms of function $\eta$.

\subsection{The triple pomeron vertex}
Corresponding to the two inhomogeneous terms in Eq. (\ref{eq41}) its
solution is split in two terms, the double pomeron exchange term,
generated by $D_{40}$, and the triple pomeron interaction term $Z\otimes D_2$.
However one can simplify the solution transferring the part of the
double pomeron exchange term leading in the high-energy limit into the
triple interaction part (~\cite{bartels, bravac}). This is achieved by
separating from the total solution the reggeized $D_{40}$ term:
\beq
D_4=D_{40}(D_{20}\to D_2)+D_4^I.
\eeq
The irreducible part $D_4^I$ proves to be a pure triple pomeron interaction,
which satisfies
\beq
D_4^I=Y\otimes D_2+(K_{12}+K_{34})D_4^I.
\eeq
The derivation again uses only the bootstrap, relation (\ref{neww})
between $W$ and $K$ and property (\ref{eta0}).
The explicit form for the new triple-pomeron vertex $Y$
turns out to be
\[
Y\otimes D_2=\frac{1}{2}\Big(G(1,23,4)+G(1,24,3)+
G(2,13,4)+G(2,14,3)+G(12,0,34)
\]\beq
-G(1,2,34)-G(2,1,34)-G(3,4,12)-G(4,3,12)
\Big)
\label{vertex}
\eeq
For a fixed coupling constant this expression was found long ago in
~\cite{bartels}. In our approach it remains true also for arbitrary
function $\eta(q)$ and thus for a running coupling introduced by means
of this function.

\section {Coupling to pomerons}
\subsection{Momentum space approach}
In the momentum space coupling the vertex $Y$ to two ougoing pomerons is
straightforward. The two pomerons are described by a product
$P(1,2)P(3,4)$ and all one is to do is to integrate this product with
$Y\otimes D_2$ over the gluon momenta 1,2,3 and 4 with 12 and 34 fixed
and $1234=0$ from the momentum conservation.

As with a fixed coupling constant, there are certain properties of the
wave function and the vertex which simplify the resulting expression.
First we expect that the pomeron wave function in the coordinate
space $P(r_1,r_2)$
vanishes if the two gluons are located at the same spatial point:
$P(r,r)=0$. This property is well-known for the BFKL pomeron with a fixed
coupling constant and is related to the behaviour of (\ref{etafix}) at
$q=0$. With a running coupling this behaviour does not change, so we
expect that the coordinate wave function will continue to vanish at $r_1=r_2$.
As a result the last five terms in the sum (\ref{vertex}) will give no
contribution, since they depend only on the sum of the momenta in one
of the pomerons and so put the two gluons at the same spatial point in it.
If the mentioned property of the pomeron wave function is violated with the
introduction of a running coupling, we still can drop the five last terms
once we restrict ourselves to the case when the two pomerons are taken
at zero total momentum: 12=34=0, which is the only case relevant for the
pomeron propagation through the nucleus. Indeed direct inspection shows
that
\beq
G(q_1,q_2,q_3)=0,\ \ {\rm if}\ \ q_1=0,\ \ {\rm or}\ \  q_3=0.
\label{gzero}
\eeq
Second, due to the symmetry of the pomeron wave function, each of the
four remaining terms in (\ref{vertex}) gives the same contribution.
So, as for the fixed coupling,  the triple pomeron vertex effectively reduces
to
\beq
Y\otimes D_2=2G(1,23,4).
\eeq

Coupling this vertex to two forward pomerons and using Eq. (\ref{defg})
we get an expression
for the triple pomeron interaction amplitude $T$ in terms of
function $\eta$ (suppressing the dependence on rapidities)
\[
T=2N_c\int \frac{d^2q_1d^2q_4d^2q'_1}{(2\pi)^6}
P(1,2)P(3,4)\]\[\Big\{-D_2(1',-1')\Big(\frac{\eta(14)}{\eta(1-1')\eta(4-4')}
-
\frac{\eta(1)}{\eta(1-1')\eta(4')}-
\frac{\eta(4)}{\eta(1')\eta(4-4')}\Big)
\]\beq
+\frac{1}{2}D_2(1,-1)\Big(\frac{\eta(14)}{\eta(1')\eta(14-1')}-
\frac{\eta(1)}{\eta(1')\eta(1-1')}\Big)
+\frac{1}{2}D_2(4,-4)\Big(\frac{\eta(14)}{\eta(1')\eta(14-1')}-
\frac{\eta(4)}{\eta(1')\eta(4-1')}\Big)\Big\}
\label{tmom}
\eeq
with $12=34=1'4'=0$.
In this formula actually both $D_2$   and $P$ are the pomeron wave functions
for the incoming and out going pomerons respectively. Function $D_2$ is the
amputated wave function related to $P$ by the relation
\beq
P(1,-1)=\frac{1}{\eta^2(1)}D_2(1,-1).
\eeq

Expression  (\ref{tmom}) is rather cumbersome. A simpler expression is
obtained in the coordinate representation, which will be presently derived.

\subsection{Coordinate space approach}
We present
\beq
(2\pi)^2\delta^2(q_{12}-q_1-q_2)P(q_1,q_2)=
\int d^2r_1d^2r_2P_{q_{12}}(r_1,r_2)e^{iq_1r_1+iq_2r_2},
\eeq
where $P_{q_{12}}(r_1,r_2)$ is the coordinate wave function of the
pomeron with the total momentum $q_{12}$.
Similarly we present the second pomeron via $P_{q_{34}}(r_3,r_4)$.
Finally
\beq
G(q_1,q_2+q_3,q_4)=\int d^2r_1d^2r_2d^2r_3e^{-iq_1r_1-i(q_2+q_3)r_2-iq_4r_3}
G(r_1,r_2,r_3).
\eeq
Then integration over $q_1,...q_4$ gives
\beq
(2\pi)^2 \delta^2(q_{12}+q_{34})T=
2\int dr^2r_1d^2r_2d^2r_3P_{q_{12}}(r_1,r_2)P_{q_{34}}(r_3,r_2)G(r_1,r_2,r_3)
\label{inieq}
\eeq
We have
\beq
P_{q_{12}}(r_1,r_2)=e^{\frac{1}{2}iq_{12}(r_1+r_2)}P(r_{12}),
\eeq
where $r_{12}=r_1-r_2$. So (\ref{inieq}) becomes
\beq
(2\pi)^2 \delta^2(q_{12}+q_{34})T=
2\int d^2r_1d^2r_2d^2r_3
e^{\frac{1}{2}i[q_{12}r_1+(q_{12}+q_{34})r_2+q_{34}r_3]}
P(r_{12})P(r_{32})G(r_1,r_2,r_3).
\label{nexteq}
\eeq
At this stage we note that if we drop $P(r_{12})$ from the integrand we
get
\[
\int dr^2r_1d^2r_2d^2r_3
e^{\frac{1}{2}i[q_{12}r_1+(q_{12}+q_{34})r_2+q_{34}r_3]}
P(r_{32})G(r_1,r_2,r_3)
\]\beq
=
\int d^2r_2d^2r_3
e^{\frac{1}{2}i[(q_{12}+q_{34})r_2+q_{34}r_3]}
P(r_{32})G(q_{12},r_2,r_3).
\label{noteeq}
\eeq
If $q_{12}=0$ then this expression vanishes due to property (\ref{gzero}).
As a result one can substitute in (\ref{nexteq})
\beq
P(r_{12})\to P(r_{12})-P(0), \ \ P(r_{32})\to P(r_{32})-P(0).
\label{subst}
\eeq
So whether $P(0)$ is equal to zero or not, for the forward case one can
always make it equal to zero by substitution (\ref{subst}). So in the
following  we assume $P(0)=0$.

As follows from the translational invariance, for the overall zero total
momentum $G(r_1,r_2,r_3)=G(r_{12},r_{32})$. So taking as integration
variables $r_2$, $r_{12}$ and $r_{32}$ we finally obtain
\beq
T=2\int d^2r_{12}d^2r_{32}P(r_{12})P(r_{32})G(r_1,r_2,r_3).
\eeq

Now we have to calculate $G(r_1,r_2,r_3)$. Due to $P(0)=0$ we may drop
all terms containing $\delta^2(r_{12})$ and/or $\delta^2(r_{23})$.
The total
contribution consists of two parts, the first one coming from
the term $-W$ in (\ref{defg}) and the second one from the rest.
As for the fixed coupling constant case (see ~\cite{bravac}) in the first
part only the first term in (\ref{wexpl}) gives a contribution which
does not contain $\delta^2(r_{12})$ nor $\delta^2(r_{23})$ nor both.
Direct calculation gives for this contribution
\beq
G_1(r_1,r_2,r_3)=-N_cD_2(r_1,r_3)\int
d^2\rho\tilde{\eta}(\rho)
\xi(r_{12}-\rho)\xi(r_{32}-\rho),
\eeq
where $\tilde{\eta}(r)$ is the Fourier transform of $\eta(q)$ and
$\xi(r)$ is the Fourier transform of $1/\eta(q)$.
From the four terms in the second part the contribution which
does not contain $\delta^2(r_{12})$ nor $\delta^2(r_{23})$ nor both
comes from the first and third terms. Its calculation gives
\beq
G_2(r_1,r_2,r_3)=\frac{1}{2}N_cD_2(r_1,r_3)\int d^2\rho
\tilde{\eta}(\rho)\Big(\xi^2(r_{12}-\rho)+\xi^2(r_{32}-\rho)\Big).
\eeq
Summing we get the final expression
\beq
G(r_1,r_2,r_3)=\frac{1}{2}N_cD_2(r_{13})F(r_{12},r_{32}),
\eeq
where
\beq
F(r_1,r_2)=\int d^2\rho
\tilde{\eta}(\rho)\Big(\xi(r_1-\rho)-\xi(r_2-\rho)\Big)^2
\label{ffunction}
\eeq
and we also used that for the forward pomeron $D_2(r_1,r_3)=D_2(r_{13})$.
The triple pomeron contribution to the amplitude is then obtained as
\beq
T(Y)=N_c\int_0^Y dy\int d^2r_{12}d^2r_{32}
F(r_{12},r_{32})P(Y-y,r_{12})P(Y-y,r_{32})D_2(y,r_{13}),
\eeq
where we restored the $y$-dependence, suppressed up to now.

Note that for a fixed coupling constant we have (see Appendix and also ~\cite{bravac})
\beq
G^{fix}(r_1,r_2,r_3)=-\frac{g^2N_c}{8\pi^3}
\frac{r_{13}^2}{r_{12}^2r_{32}^2}\Big(g^2 D_2(r_{13})\Big).
\label{gfixcoup}
\eeq
We can consider our new expression as a result of changing the fixed
$g^2$ to a running quantity
\beq
g^2\to -\frac{4\pi^3r_{12}^2r_{32}^2}{r_{13}^2}F(r_{12},r_{32}).
\label{runcoup}
\eeq

Function $F(r_1,r_2)$ can be presented in a different form, which demonstrates
absence of ultraviolet divergency coming from the singular behavior of $\tilde{\eta}(\rho)$
at $\rho\to 0$. We have
\[
\int d^2\rho
\tilde{\eta}(\rho)\xi^2(r_{1}-\rho)=\int d^2\rho\frac{d^2q}{(2\pi)^2}\frac{d^2q_1}{(2\pi)^2}
\frac{d^2q_2}{(2\pi)^2}\frac{\eta(q)}{\eta(q_1)\eta(q_2)}e^{iq\rho+iq_1(r_1-\rho)+iq_2(r_1-\rho)}
\]\beq
=\int\frac{d^2q_1}{(2\pi)^2}\frac{d^2q_2}{(2\pi)^2}\frac{\eta(q_1+q_2)}{\eta(q_1)\eta(q_2)}
e^{ir_1(q_1+q_2)}
\eeq
and similarly
\[
\int d^2\rho
\tilde{\eta}(\rho)\xi(r_{1}-\rho)\xi(r_2-\rho)
=\int d^2\rho\frac{d^2q}{(2\pi)^2}\frac{d^2q_1}{(2\pi)^2}
\frac{d^2q_2}{(2\pi)^2}\frac{\eta(q)}{\eta(q_1)\eta(q_2)}e^{iq\rho+iq_1(r_1-\rho)+iq_2(r_2-\rho)}
\]\beq
=\int\frac{d^2q_1}{(2\pi)^2}\frac{d^2q_2}{(2\pi)^2}\frac{\eta(q_1+q_2)}{\eta(q_1)\eta(q_2)}
e^{ir_1q_1+ir_2q_2}.
\eeq
So we find
\beq
F(r_1,r_2)=\int\frac{d^2q_1}{(2\pi)^2}\frac{d^2q_2}{(2\pi)^2}\frac{\eta(q_1+q_2)}
{\eta(q_1)\eta(q_2)}
\Big(e^{iq_1r_1}-e^{iq_1r_2}\Big)\Big(e^{iq_2r_1}-e^{iq_2r_2}\Big).
\eeq
In this form it is clear that $F(r_1,r_2)$ is a well defined function which does not contain
ultraviolet nor infrared divergency.

For further use note the identity
\beq
\int d^2r_1F(r_1-r,r_1)=\int\frac{d^2q_1}{(2\pi)^2}\frac{d^2q_2}{(2\pi)^2}\frac{\eta(q_1+q_2)}
{\eta(q_1)\eta(q_2)}
\Big(e^{-iq_1r}-1\Big)\Big(e^{-iq_2r}-1\Big)\int d^2r_1e^{ir_1(q_1+q_2)}=0,
\label{iden}
\eeq
since $\eta(0)=0$.
%%%%%%%%%%%%%%%%%%%%%%%%%%%%%%%%%%%%%%%%%%%%%%%%%%%%%%%%%%%%%%%%%%%%%%%%%%%%%%%%%%%%%%%%%%%%%%%
%%%%%%%%%%%%%%%%%%%%%%%%%%%%%%%%%%%%%%%%%%%%%%%%%%%%%%%%%%%%%%%%%%%%%%%%%%%%%%%%%%%%%%%%%%%%%%

\section{Pomeron in the nucleus}
With the triple Pomeron vertex known, it is straightforward to obtain the equation which sums fan
diagrams  describing propagation of the pomeron in the nucleus. Repeating the derivation for the
fixed coupling constant in ~\cite{braun4} we find for this sum $\Phi(y,b,r)$ at fixed impact parameter
$b$:
\[
\Phi(y,r,b)=\Phi_1(y,b,r)+\frac{1}{2}N_c\int_0^\infty dy'\prod_{j=1}^3d^2r_j
\delta^2(r_1-r_2+r_3)
\]\beq
F(r_2,r_3)\eta^2(-i\nabla_1)G(y-y',r,r_1)\Phi(y',r_2,b)\Phi(y',r_3,b).
\label{fan}
\eeq
Here $G(y,r,r')$ is the pomeron forward Green function satisfying the equation
\beq
 \Big(\frac{\pd}{\pd y}+H\Big)G(y,r,r')=\delta(y)\eta^{-1}(-i\nabla)\eta^{-1}(-i\nabla')
\delta^2(r-r'),
\eeq
with the Hamiltonian $H$ for the non-amputated forward wave function given  by
\beq
H=2\omega+K^{\dagger},
\label{ham}
\eeq
where $\omega$ is expressed  via function $\eta$ according to (\ref{traj}) and
\beq
K^{\dagger}(q_1|q'_1)=2N_c\frac{\eta(q'_1)}{\eta(q_1)\eta(q_1-q'_1)}.
\label{forint}
\eeq
The inhomogeneous term $\Phi_1$ corresponds to the single pomeron exchange:
\beq
\Phi_1(y,r,b)=\frac{1}{2}N_cAT(b)\int d^2r'G(y,r,r')\rho(r'),
\eeq
where $T(b)$ is the nuclear profile function and $\rho(r)$ is the color density of the nucleon.
Applying   operator $\pd/\pd y+H$ to Eq. (\ref{fan}) we find the evolution equation for the
pomeron in the nucleus
\beq
\Big(\frac{\pd}{\pd y}+H\Big)\Phi(y,r,b)=\delta(y)\Phi_0(r,b)+
\frac{1}{2}N_c\int\prod_{j=2}^3d^2r_j\delta^2(r-r_2+r_3)
F(r_2,r_3)\Phi(y,r_2,b)\Phi(y,r_3,b),
\label{faneq}
\eeq
where $\Phi_0(r,b)$, playing the role of the initial condition, is given by
\beq
\Phi_0(r,b)=\frac{1}{2}AT(b)\eta^{-2}(-i\nabla)\rho(r).
\label{iniphi}
\eeq

Unlike the case of the fixed coupling constant this equation is not
simplified in the momentum space.

To compare with the dipole approach, we rewrite
our Eq. (\ref{faneq}) as a whole in the coordinate space. To do this we
have to rewrite action of the Hamiltonian (\ref{ham}) on the wave function in
the coordinate space. In transforming the pomeron amplitude $\Phi(y,q,b)$ to
the coordinate space we have to take into account condition
\beq
\Phi(y,r=0,b)=0
\label{phieq0}
\eeq
which we have extensively used. Technically it means that we have to add
to $\Phi(y,q,b)$ a term proportional to $\delta^2(q)$ which guarantees this
property. This  is essential to obtain the correct form for
the linear part of the evolution equation in the coordinate space
\footnote{We are greataly indebted to Yu. Kovchegov who pointed out this
circumstance.}

It is  convenient to split (\ref{ffunction}) into three terms
\beq
F(r_1,r_2))=f(r_1,r_1)+f(r_2,r_2)-2f(r_1,r_2),
\label{sumf}
\eeq
where
\beq
f(r_1,r_2)=\int d^2\rho
\tilde{\eta}(\rho)\xi(r_1-\rho)\xi(r_2-\rho).
\label{ffun1}
\eeq
In terms of this function one easily finds
\beq
2\int \frac{d^2q}{(2\pi)^2}e^{iqr}\omega(q)\Phi(y,q,b)=
N_c\int d^2r_1f(r_1-r,r_1-r)\Phi(y,r_1,b)
\eeq
and
\beq
\int \frac{d^2q}{(2\pi)^2}\frac{d^2q'}{(2\pi)^2}e^{iqr}K^{\dagger}(q|q')\Phi(y,q',b)=
2N_c\int d^2r_1f(r_1-r,r_1)\Phi(y,r_1,b).
\eeq
%%%%%%%%%%%%%%%%%%%%%%%%%%%%%%%%%%%%%%%%%%%%%%%%%
Thus in the coordinate space we get
\beq
H\Phi(y,r,b)=-N_c\int d^2r_1\Big(2f(r_1-r,r_1)-f(r_1-r,r_1-r)\Big)
\Phi(y,r_1)+\ \ const,
\eeq
where $const$ should be taken to ensure property (\ref{phieq0}).
As a result we find
\beq
H\Phi(y,r,b)=N_c\int d^2r_1F(r_1-r,r_1)\Phi(y,r_1,b)=
\frac{1}{2}N_c\int d^2r_1F(r_1-r,r_1)\Big(\Phi(y,r_1,b)+\Phi(y,r_1-r,b)\Big).
\eeq
Using identity (\ref{iden}) we may add to the bracket any function independent of $r_1$,
to finally obtain
\beq
H\Phi(y,r,b)=
\frac{1}{2}N_c\int d^2r_1F(r_1-r,r_1)\Big(\Phi(y,r_1,b)+\Phi(y,r_1-r,b)-\Phi(y,r,b)\Big).
\eeq
In this form the linear part of the evolution equation avquires the standard colour
dipole structure (see e.g. ~\cite{kovchegov2}) and the whole evolution equation
becomes
\[
\frac{\pd}{\pd y}\Phi(y,r)=-\frac{1}{2}N_c\int d^2r_1F(r_1-r,r_1)\]\beq
\Big(\Phi(y,r_1,b)+\Phi(y,r_1-r,b)-\Phi(y,r,b)-
\Phi(y,r_1,b)\Phi(y,r_1-r,b)\Big).
\label{forintr}
\eeq

In the limit of the fixed coupling constant
we find, dropping the infrared regularization terms,
\beq
f^{fix}(r_1,r_2)=-\frac{\alpha_s}{\pi^2}\frac{{\bf r}_1{\bf r}_2}{r_1^2 r_2^2},
\eeq
where at $r_1=r_2=r$ one should understand $1/r^2$ as regularized in the ultraviolet
(see ~\cite{bravac}):
\beq
\frac{1}{r^2}\equiv\frac{1}{r^2+\epsilon^2}+2\pi\delta^2(r)\ln\epsilon,\ \ \epsilon\to 0.
\eeq
So in analogy with ~\cite{kovchegov1} we may define 3 running coupling constants by
\beq
f(r_1,r_2)=-\frac{1}{\pi^2}\frac{\alpha_s(r_1)\alpha_s(r_2)}{\alpha_s(r_1,r_2)}
\frac{{\bf r}_1{\bf r}_2}{r_1^2r_2^2},
\label{runco}
\eeq
with the additional condition $\alpha_s(r,r)=\alpha_s(r)$,
and rewrite Eq. (\ref{forintr}) as
\[
\frac{\pd}{\pd y}\Phi(y,r,b)=
\frac{1}{2\pi^2}N_c\int d^2r_2d^2r_3\delta(r-r_1+r_2)F(r_1,r_2)
\Big(\frac{\alpha_s(r_1)}{r_1^2}+\frac{\alpha_s(r_2)}{r_2^2}
-2\frac{\alpha_s(r_1)\alpha_s(r_2)}{\alpha_s(r_1,r_2)}
\frac{{\bf r}_1{\bf r}_2}{r_1^2r_2^2}\Big)\]\beq
\Big(\Phi(y,r_1,b)+\Phi(y,r_2,b)-\Phi(y,r,b)-
\Phi(y,r_1,b)\Phi(y,r_2,b)\Big).
\label{forintr2}
\eeq

\section{Discussion}

Eqs. (\ref{forintr}) and  (\ref{forintr2}) present our final result
for the non-linear BFKL equation with the running coupling. We
stress that function $\eta(q)$ in them is determined only by its
asymptotic form (\ref{asym}) together with the requirement
(\ref{eta0}). A simple possibility is to choose
\beq
\eta(q)=\frac{1}{2\pi}bq^2\ln\Big(a+\frac{q^2}{\Lambda^2}\Big),
\eeq
with $b$ given by (\ref{bval}) and arbitrary $a>1$.
Also one has to remember that the equations are derived only in the leading order
in the running coupling.
Already subleading terms of the relative order $1/\ln(q^2/\Lambda^2)$ remain
undetermined, since they correspond to the next-to-leading order in the running coupling.

Our final coordinate space equation (\ref{forintr2}) fully coincides with Eq. (101)
in ~\cite{kovchegov1}
obtained in the dipole formalism (for the forward case).
However in our approach the running couplings $\alpha_s(r)$ and $\alpha_s(r_1,r_2)$
are defined by  Eq. (\ref{runco}) in a general manner, irrespective to
any regularization procedure and are determined by the concrete choice of function $\eta(q)$.
In fact they are fixed only in as far  the high-momentum behaviour of this function is
known and so admit a high degree of arbitrariness. It remains to be seen how this
arbitrariness influence concrete results which follow from the solution of the
evolution equation (\ref{forintr2}).

\section{Appendix. Function $F(r_1,r_2)$ with a fixed coupling constant}
We check that for a fixed coupling constant function $F(r_1,r_2)$ indeed passes into the expression corresponding to
Eq. (\ref{gfixcoup}). In the fixed coupling case $\eta(k)$ is given by (\ref{etafix}) Then we find
\beq
\tilde{\eta}(\rho)=-\frac{2\pi}{g^2}\nabla^2\delta^2(\rho)
\eeq
and
\beq
\xi(r)=-\frac{g^2}{(2\pi)^2}(\ln r-c),
\eeq
where $c=\ln(2/m)+\psi(1)$ and $m$ is the infrared regularizer (gluon mass).
Terms containing $c$ and thus depending on the infrared regularization cancel  in the
final result.
Performing the integration over $\rho$ with the help of the $\delta$-function we find
\beq
f^{fix}(r_1,r_2)= -\frac{ g^2}{(2\pi)^3}\Big((\ln r_1-c)\nabla_2^2\ln r_2+(\ln
r_2-c)\nabla_2^2\ln r_1+
2\nabla_1\ln r_1\nabla_2\ln r_2\Big).
\eeq
Taking into account that
\beq
\nabla^2\ln r=2\pi \delta^2(r),
\eeq
we find
\beq
f^{fix}=-\frac{g^2}{8\pi^3}\Big(2\frac{{\bf r}_1{\bf r}_2)^2}{r_1^2r_2^2}
+2\pi\delta^2(r_1)(\ln r_2-c) +2\pi\delta^2(r_2)(\ln r_1-c)\Big).
\eeq
Forming $F(r_1,r_2)$ according to (\ref{sumf}) we see that terms with $c$ cancel
and dropping terms proportional to $\delta^2(r_1)$ or $\delta^2(r_2)$ we find
\beq
F^{fix}(r_1,r_2)=-\frac{g^2}{4\pi^3}\frac{(r_1-r_2)^2}{r_1^2r_2^2},
\eeq
in full correspondence with (\ref{gfixcoup}).

\section{Acknowledgements}
The author would like to thank Yu. Kovchegov for numerous informative
discussions. This work has been supported by grants RNP 2.1.1.1112
and RFFI 06-02-16115a of Russia.


\begin{thebibliography}{99}
%
\bibitem{kovchegov1} Yu.V.Kovchegov and H.Weigert, hep-ph/0609090.
%
\bibitem{kovchegov2}  Yu.V.Kovchegov and H.Weigert, hep-ph/0612071.
%
\bibitem{balitsky} I.I.Balitsky, hep-ph/0609087.
%
\bibitem{braun1} M.A.Braun, Phys.Lett. {\bf B 345} (1995) 155.
%
\bibitem{braun2} M.A.Braun, Phys.Lett. {\bf B 348} (1995) 190.
%
\bibitem{bartels0} J.Bartels, Nucl. Phys, {\bf B 175} (1980) 365.
%
\bibitem{braun3} M.A.Braun, Eur. Phys. J {\bf C 6} (1999) 321.
%
\bibitem{bravac} M.A.Braun and G.P.Vacca, Eur. Phys. J. {\bf C 6} (1999) 147.
%
\bibitem{bartels} J.Bartels and M.Wuesthoff, Z.Phys. {\bf C 66} (1995) 157.
%
\bibitem{braun4} M.A.Braun, Eur. Phys. J. {\bf C 16} (2000) 337.
%
\end{thebibliography}
\end{document}